\documentclass[prd,10pt,aps,showpacs,twocolumn,longbibliography,nofootinbib,amsmath,amssymb,superscriptaddress,longbibliography]{revtex4-1}

\usepackage{graphicx}  
\usepackage{subfigure}
\usepackage{multirow}
\usepackage{breqn}
\usepackage{longtable}
\usepackage[T1]{fontenc}
\usepackage{dcolumn}   
\usepackage{bm}        
\usepackage{amsfonts}  
\usepackage{amsmath}   
\usepackage{amssymb}   
\usepackage{gensymb}
\usepackage[colorlinks=True, linkcolor=violet, citecolor=teal, urlcolor=blue]{hyperref}
\usepackage{xcolor}

\usepackage[english]{babel} 

\graphicspath{{figures/}}

\newcommand{\pwisein}{\left\{ \begin{array}{ll}}
\newcommand{\pwiseout}{\end{array}\right.}

\newcommand{\beq}{\begin{equation}}
\newcommand{\eeq}{\end{equation}}
\newcommand{\beqn}{\begin{eqnarray}}
\newcommand{\eeqn}{\end{eqnarray}}

\allowdisplaybreaks

\begingroup\makeatletter
\catcode`,=\active
\global\let\breqn@comma,
\protected\gdef,{\ifmmode\expandafter\breqn@comma\else\expandafter\active@comma\fi}
\endgroup

\begin{document}
\bibliographystyle{apsrev4-1}
\title{The Effect of a Knot on the Thermal Stability of Protein MJ0366 : Insights from Molecular Dynamics and Monte Carlo Simulations}

\author{A. M. Begun}
\affiliation {Pacific Quantum Center, Far Eastern Federal University, 690922, Vladivostok, Russia}
\affiliation {Nordita, Stockholm University, Stockholm, Sweden}
\author{A. A. Korneev}
\affiliation {Pacific Quantum Center, Far Eastern Federal University, 690922, Vladivostok, Russia}
\author{Anastasiia Zorina}
\affiliation {Pacific Quantum Center, Far Eastern Federal University, 690922, Vladivostok, Russia}

\begin{abstract}
    Protein MJ0366  is a hypothetical protein from Methanocaldococcus jannaschii that has a rare and complex knot in its structure. The knot is a right-handed trefoil knot that involves about half of the protein's residues. In this article, we investigate the thermal stability of protein MJ0366  using numerical simulations based on molecular dynamics and Monte Carlo methods. We compare the results with those of a similar unknotted protein and analyze the effects of the knot on the folding and unfolding processes. We show that the knot in protein MJ0366  increases its thermal stability by creating a topological barrier that prevents the protein from unfolding at high temperatures. We also discuss the possible biological implications of the knot for the function and evolution of protein MJ0366 .
\end{abstract}

\maketitle

\section{Introduction}

\begin{figure}[ht]
    \centering
    \includegraphics[width=0.9\linewidth]{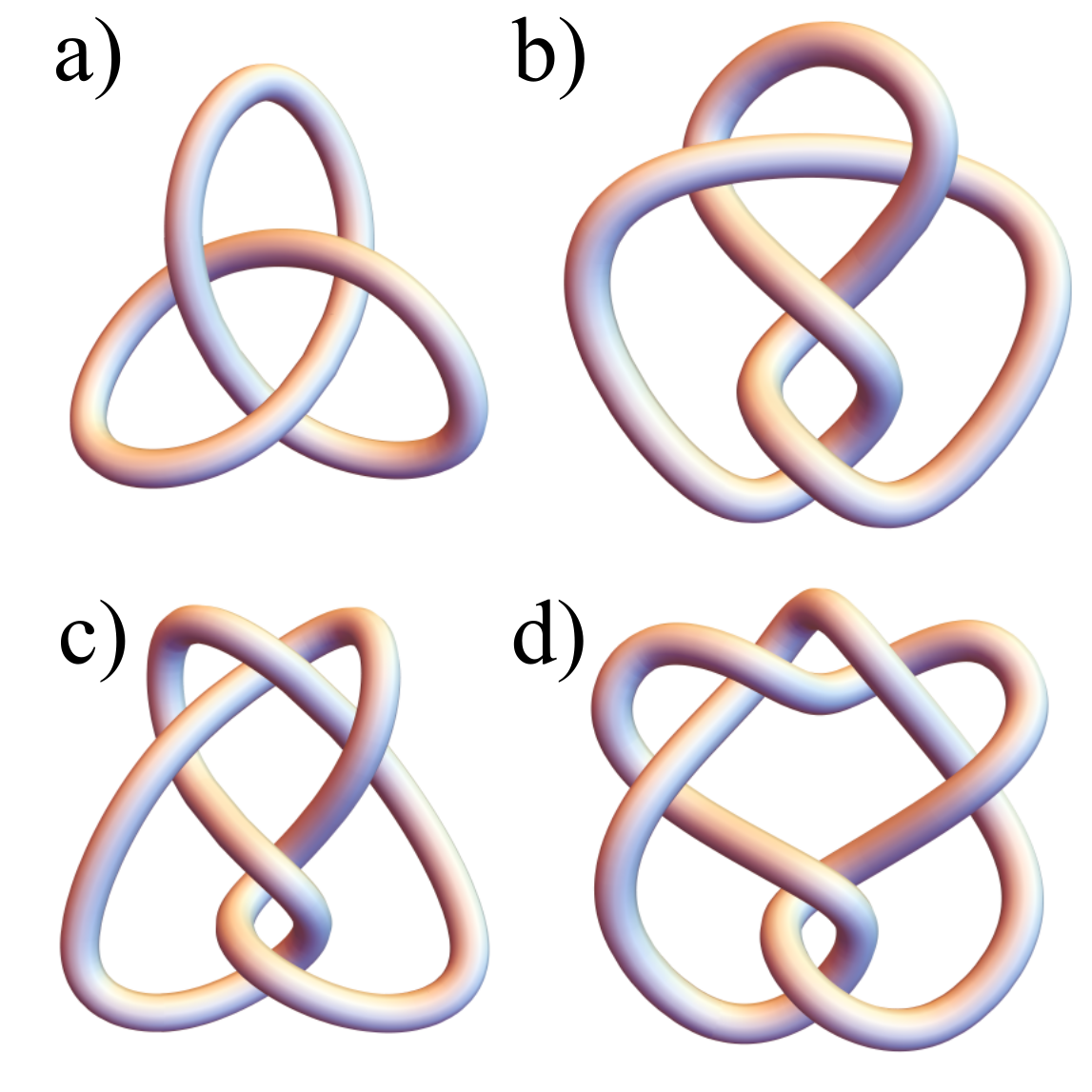}
    \caption{Possible knot types. Panel a) stands for trefoil ($3_1$) knot, panel b) is figure eight ($4_1$) knot, panel c) three-twist ($5_2$) knot, d) Stevedor knot ($6_1)$}
    \label{fig:knot-types}
\end{figure}

About 1 \% of protein molecule structures presented in the PDB bank have a knotted structure~\cite{Jamroz2014, Dabrowski-Tumanski2018}. Their backbone can be tied into different types of the simplest knots (see examples in Fig.~\ref{fig:knot-types}) according to the list of tabulating prime knots~\cite{Hoste1998}. In a protein, these knots are interpreted as topological barriers that dramatically affect the protein folding processes. In particular, the presence of such a topological barrier affects the temperature stability of the molecule~\cite{Sulkowska2020, Hsu2023}.
Moreover, knotted methyltransferases exhibit a trefoil configuration even in the denatured state~\cite{Mallam2010, Wang2013}, which also indicates a non-trivial role for the knot in protein molecules.

The hypothetical protein MJ0366  from \textit{Methanocaldococcus jannaschii} (PDB ID -  2EFV) has a knot structure of a left-handed trefoil and is the shortest knot protein available in the PDB bank~\cite{Thiruselvam2017}. Due to its small size (82 amino acids) and single-domain structure, it is the subject of many theoretical~\cite{Noel2010, Beccara2013, Noel2013, Chwastyk2015, Dabrowski-Tumanski2015, Soler2016, Niewieczerzal2017, Paissoni2021, Especial2021, Wang2022} 
and experimental studies~\cite{Wang2015, Ramirez2017, Rivera2020, Rivera2023}. One of the first experiments was the study of the folding dynamics and kinetic pathways of MJ0366 under chemical denaturation~\cite{Wang2015}. Then single molecule force spectroscopy experiment was showed that Immunoglobulin Binding Protein affects on the folding of MJ0366~\cite{Ramirez2017}. Using optical tweezers, it was shown that the mechanical folding process exhibits a two-state mechanism~\cite{Rivera2020}. Moreover, the results of experiments on the study of the temperature stability of this protein were recently presented, where a non-trivial temperature dependence was observed~\cite{Rivera2023}. 

At the same time, the conducted theoretical studies lead to different results. In this connection, the question arises about the correct description of folding/unfolding of this MJ0366 protein.

One of the first paper on \textit{in silico} study of MJ0366 used all-atom structure-based model~\cite{Noel2010}. Knot formation was shown to occur in three stages (with an intermediate slipknot conformation) begins with the $C$-end of the chain being threaded through the twisted loop. Later, a full-atom model , but with realistic interatomic forces based on DPR approach, showed qualitatively the same results~\cite{Beccara2013}. It was demonstrated that open region of that twisted loop created by native $\beta$-sheet. In~\cite{Noel2013} analogous spontaneous knotting has been demonstrated using unbiased all-atom explicit-solvent simulations. Subsequently, in a coarse-grained model, it was shown that such a single-loop formation is not dominant for MJ0366, and a more frequent number of folding trajectories fall on the two-loop mechanism~\cite{Chwastyk2015}. Simulations in different approaches were also carried out on the influence of chaperones on knotting processes~\cite{Soler2016, Niewieczerzal2017}.

Determination of the minimum number of amino acid contacts required to form a knot~\cite{Dabrowski-Tumanski2015} and a method for identifying key residues in a knot~\cite{Wang2022} were given using the MJ0366 as an example. Folding stability of this protein has also been theoretically studied by by mutating individual residues~\cite{Paissoni2021}. The set of heterogeneous native residues required for efficient knotting was also investigated~\cite{Especial2021}.

In order to study the dynamic properties of MJ0366, we use the Landau-Ginzburg-Wilson approach, which is based on the principles of universality and gauge symmetry~\cite{Chernodub2010}.
This approach has already been used to study biophysical properties of proteins. A phase diagram of myoglobin was constructed~\cite{Begun2019}. Investigation of the folding of a protein that has a slipknot type of a knot in its native state~\cite{Begun2021}. The temperature stability of $\alpha$-synuclein, which is one of the components of Lewy bodies and a pathogenic criterion for various neuro diseases, in particular, Parkinson's disease, was also assessed~\cite{Korneev2022}.
In this paper we study the folding pathways of MJ0366 by using the Landau-Ginzburg-Wilson approach. We obtain tertiary structure of the protein in terms of kinks~\cite{Chernodub2010} and analyze how the knot affects on the temperature stability of the protein MJ0366.

\section{Materials and methods}

\subsection{The model}

We use the approach of topological solitons developed by Antti J. Niemi \cite{Niemi_2014}. It is based on representing the protein as a partially linear chain consisting of $C_\alpha$ atoms and links that connect them (backbone). The positions of all $C_\alpha$ atoms in the protein determine the positions of all other atoms due to rigidity of the peptide group.

\begin{figure}
    \centering
    \includegraphics[width=1\linewidth]{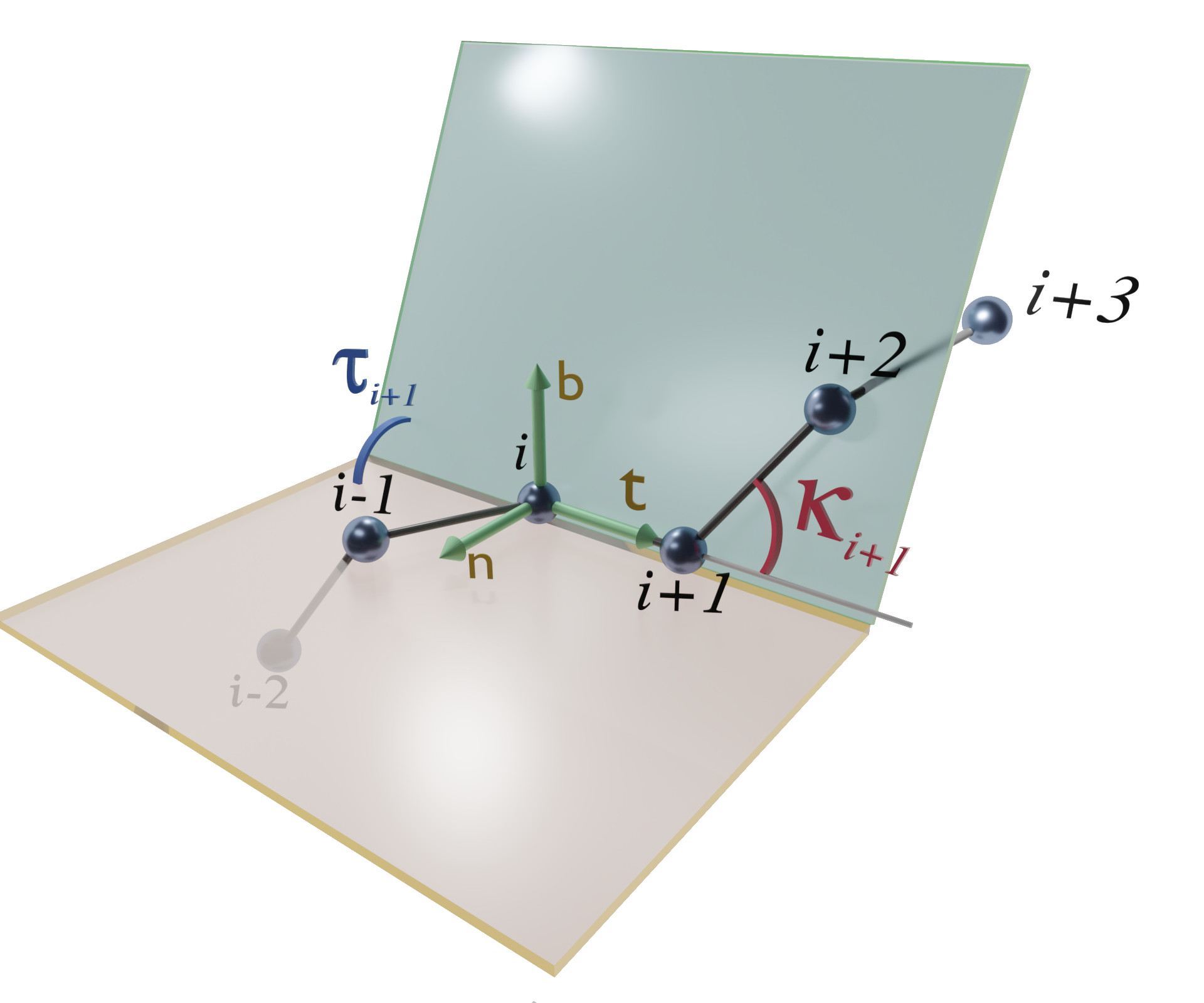}
    \caption{Backbone and Frenet frame $(\Vec{t}, \Vec{b}, \Vec{n})$}
    \label{fig:backbone}
\end{figure}

This curve is considered in terms of discrete Frenet coordinates. In each residue, the Frenet frame (Fig. \ref{fig:backbone}) consisting of three vectors $(\Vec{t}, \Vec{b}, \Vec{n})$ is built:

\begin{dgroup}
\begin{dmath}
\vec t_i=\frac{\vec r_{i+1}-\vec r_i}{|\vec r_{i+1}-\vec r_i|},
\end{dmath}
\begin{dmath}
\Vec b_i=\frac{[\vec t_{i-1},\vec t_{i}]}{|[\vec t_{i-1},\vec t_{i}]|},
\end{dmath}
\begin{dmath}
\Vec n_i=[\vec b_i, \vec t_i],
\end{dmath}
\end{dgroup}

where $\vec r_i$ are coordinate vectors of $C_\alpha$ atoms.

The angles of curvature $\kappa_i$ and torsion $\tau_i$ (Fig. \ref{fig:backbone}) act as new coordinates. They can be calculated as follows:
\begin{dgroup}[noalign]
\begin{dmath}
\kappa_i=\arccos(\vec t_{i+1}, \vec t_i),
\end{dmath}
\begin{dmath}
\tau_i =\mathrm{sign}([\vec b_{i-1},\vec b_i], \vec t_i)\arccos(\vec b_{i+1}, \vec b_i).
\end{dmath}
\end{dgroup}

The free energy function is based on the Abelian Higgs model.

\begin{dmath}
F=\sum_{i=1}^{N-1}\left(\kappa_{i+1}-\kappa_i\right)^2+\sum_{i=1}^N\left(\lambda(\kappa_i^2-m^2)^2+d\kappa_i^2\tau_i^2\\
-b\kappa_i^2\tau_i-a\tau_i+c\tau_i^2\right)+\sum_{i>j}^N V(|\vec r_i- \vec r_j|), \label{eq:energyf}   
\end{dmath}

where the terms with coefficients $a$ and $b$ stand for conservation of charges in discrete non-linear Schrodinger equation hierarchy called helicity and momentum, this terms provide parity breaking and make protein right-handed chiral. Terms with coeficients $c$ and $d$ together constitutes the Kirchhoff energy of an elastic rod. The last term is hard-core Pauli repulsion to avoid self-crossings.

This energy function possesses solutions in the form of topological solitons, and the protein model is built as a composition of solitons (multisoliton). Each soliton describes a region of the protein consisting of parts of regular structures ($\alpha$-helices and $\beta$-strands) and an intermediate structure between them (loop). 

The sets of coefficients $(\lambda, m,a,b,c,d)_k$ are individual and uniform for each soliton. Based on these coefficients, the native state can be calculated as a configuration satisfying to the minimum of \eqref{eq:energyf}:

\begin{dgroup}
\begin{dmath}
    \frac{\partial F}{\partial \kappa_i}=2(2\kappa_i-\kappa_{i-1}-\kappa_{i+1})+4\kappa_i(\kappa_i^2-m^2) +2d\kappa_i\tau_i^2{-2b\kappa_i\tau_i=0,}
\end{dmath}
\begin{dmath}
    \frac{\partial F}{\partial \tau}={2d\kappa_i^2\tau_i-b\kappa_i^2-a+2c\tau_i=0.}
\end{dmath}
\end{dgroup}

Taking into account that solitons usually aggregate several amino acid residues, the number of parameters needed to describe the protein structure is often less than the amount of amino acids that make up the protein.

As a starting point for modeling the protein structure, we use a structure taken from RCSB \cite{Berman_2000}. Using a simulated annealing  algorithm \cite{Pincus_1970}, the soliton parameters are chosen such that the configuration corresponding to the minimum free energy corresponds to the initial structure taken from the RCSB data base with a minimum RMSD value.

\subsection{Unfolding-folding simulation}

The obtained structure can be used for simulation of the unfolding-folding process using Monte Carlo methods. To do this we use Glauber algorithm \cite{Glauber_1963,Bortz_1975}. In comparison to Metropolis-Hastings algorithm it allows us to choose perturbing site randomly regardless if it was chosen before. 

At each Monte Carlo step $n$ we independently change either the torsion angle
or the curvature angle at a randomly chosen site $i$:

\begin{dgroup}
    \begin{dmath}
        \tau_i\rightarrow \tau_i+r\sigma_\tau,       
    \end{dmath}
    \begin{dmath}
        \kappa_i\rightarrow \kappa_i+r\sigma_\kappa,       
    \end{dmath}
\end{dgroup}
where $r$ is a normally distributed around $r=0$ random number with dispersion $\Delta r=1$ and $\sigma_\tau=\sigma_\kappa=0.01$ is proportionality factor. And then compute the corresponding change $\Delta F(\kappa_i,\tau_i)$ in the free energy. The obtained configuration is being accepted with probability

\beq
P=\frac{1}{1+e^{\Delta F/T_G}},
\label{eq:probability}
\eeq

where $T_G$ is the Glauber temperature factor that allows to simulate temperature changing.  It is important to note that it is Monte Carlo temperature and it can not be concided with the physical temperature measured in Kelvin directly. The relation can be found by the methods of renormalization \cite{Krokhotin2013,Peng2016}.

It has been demonstrated \cite{Martinelli1994,Martinelli1994a} that Monte Carlo simulation using equation \eqref{eq:probability} converges exponentially to the Gibbsian distribution. Therefore, this approach should offer a statistically significant representation of protein folding dynamics near equilibrium during adiabatic temperature changing.

The simulation process consists of three stages: heating (unfolding), thermalization, cooling (folding). During heating and cooling temperature is increasing (decreasing) by exponential law with growth factor defined by required amount of steps and temperature range:
\beq
T_G'=T_Gb,
\eeq
\beq
b=ln\frac{T_{max}}{T_{min}}N^{-1},
\eeq
where $N$ is number of steps required for each stage. This choice allows to guarantee adiabatic nature of the process.

\subsection{Knot detection}

Knots as topological objects are defined only on closed loops while proteins mostly are not closed. To avoid this different techniques can be applied. 

\begin{figure}[ht]
    \centering
    \includegraphics[width=1\linewidth]{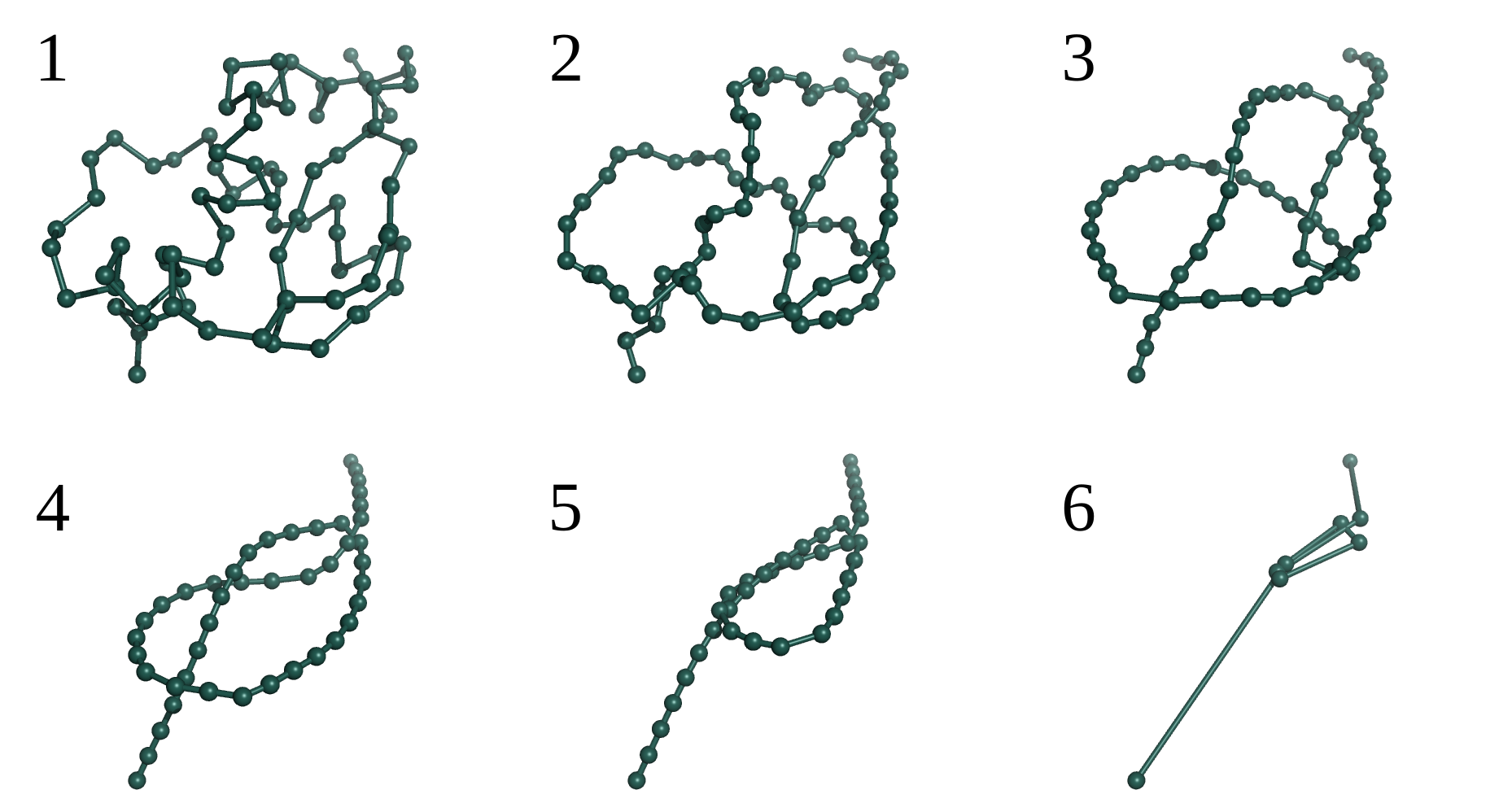}
    \caption{Sequential states during the Knot\_pull calculations of 2EFV knot type.}
    \label{fig:knot_pull}
\end{figure}

To detect whether protein is knotted or not we used Knot\_pull—python package \cite{Jarmolinska2019} by Aleksandra Jarmolińska. Knot\_pull refines and streamlines the given structure with topological awareness, ensuring that the inherent chain does not intersect with itself (Fig. \ref{fig:knot_pull}). Smoothing continues until no further modifications are possible (or the current arrangement has been previously examined), any unnecessary beads are eliminated, retaining only the essential topological points. Based on this structure, the corresponding Alexander-Briggs \cite{Alexander1926} notation is calculated.

\begin{figure}[ht]
    \centering
    \includegraphics[width=0.8\linewidth]{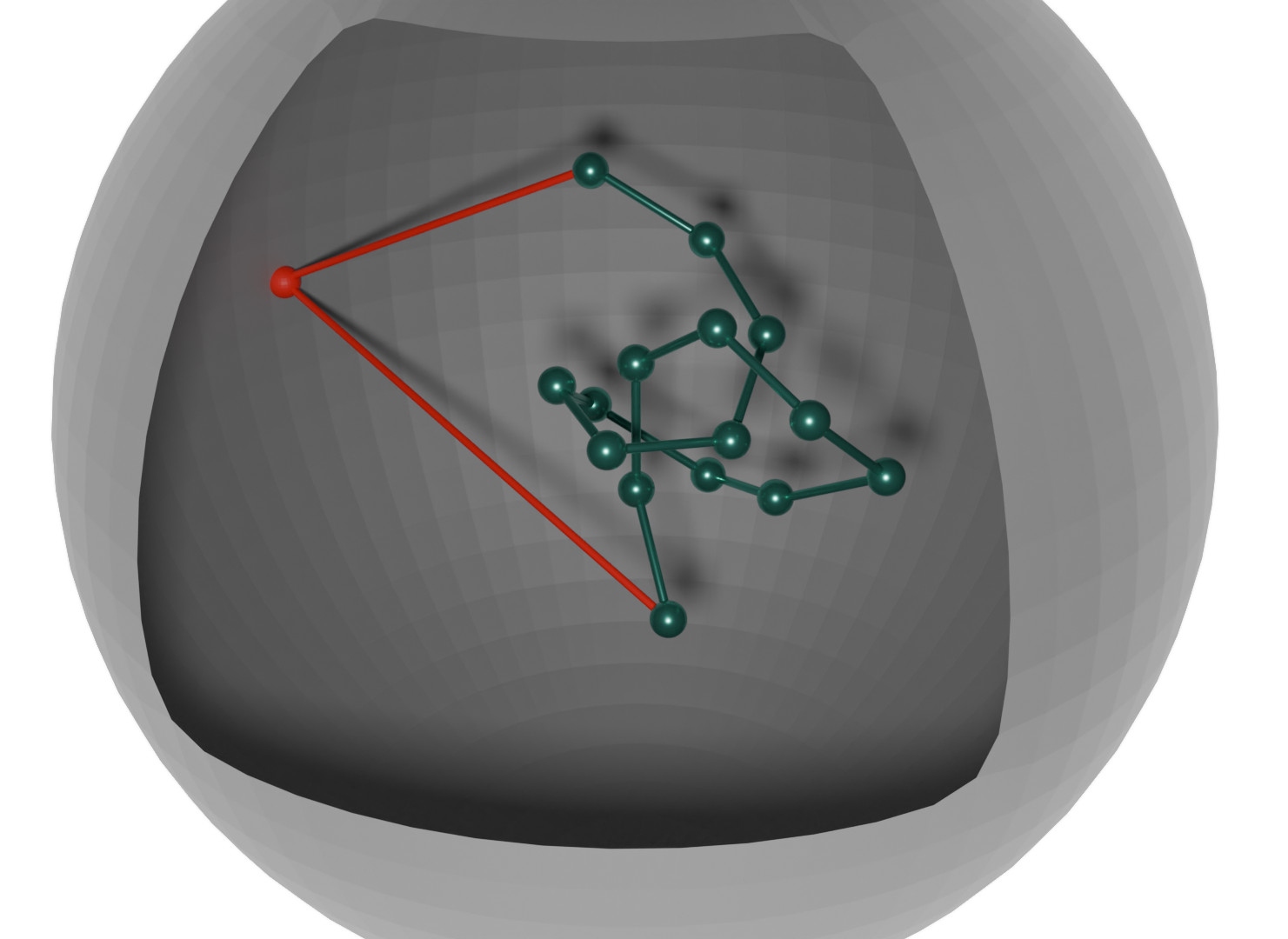}
    \caption{Closure the protein structure on infinite sphere. By red color additional bounds are marked.}
    \label{fig:closure}
\end{figure}

This approach is more suitable than others based on the calculation of polynomial invariants for the determination of knot type \cite{Lua2012, Tubiana2018, DabrowskiTumanski2018,DabrowskiTumanski2020}. To detect knot type these approaches close the loop randomly on infinite radius sphere (Fig. \ref{fig:closure}), such closure repeats many times and gives only the probability of observing given knot type. 

\subsection{Studied protein}

The protein MJ0366 was selected for investigation due to its status as the smallest known knotted protein. Its relatively small size and knotted topology make it a particularly interesting subject for studying the structural and functional implications of protein knots, providing a simpler model to understand the biological significance and stability of these complex protein structures.
 
The MJ0366 protein is a hypothetical protein from the archaeon \textit{Methanocaldococcus jannaschii} \cite{Thiruselvam2017}, which is a hyperthermophilic, methane-producing microorganism found in extreme environments such as hydrothermal vents.

\textit{M. jannaschii} is a member of the Archaea domain, a group of prokaryotes that often thrive in extreme environments. It grows optimally at temperatures above $85 \celsius$ and can reduce carbon dioxide to methane using hydrogen, a process known as methanogenesis.
This species is used as a model organism to study extremophilic life and early evolutionary processes due to its ancient evolutionary lineage and unique biochemical pathways.

\begin{figure}[ht]
    \centering
    \begin{tabular}{c c}
    \includegraphics[width=0.45\linewidth]{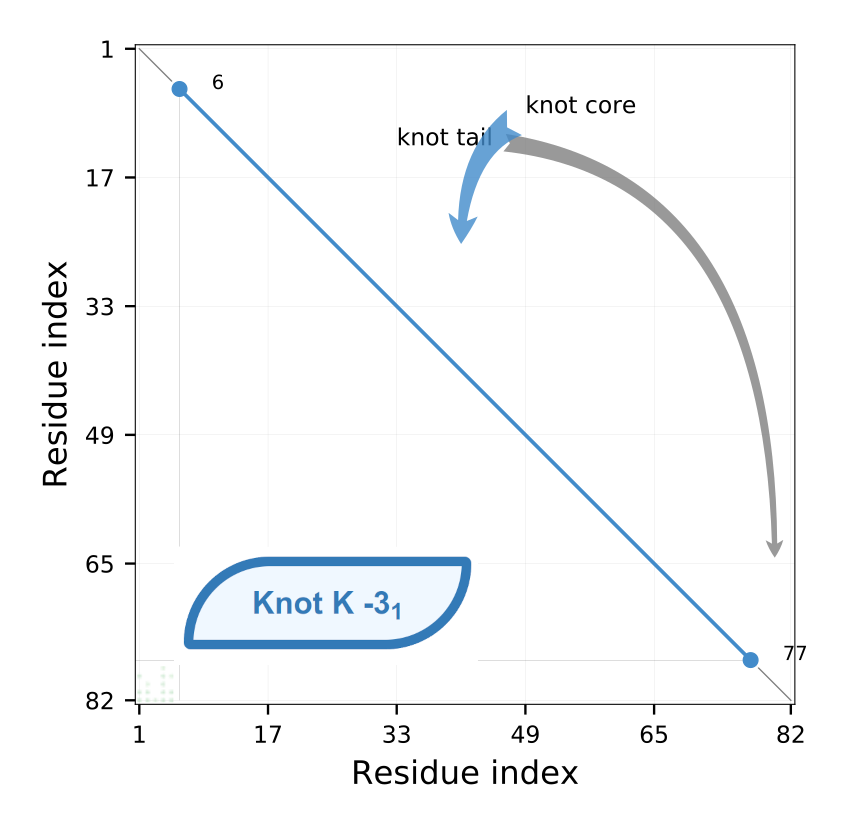}     &      \includegraphics[width=0.45\linewidth]{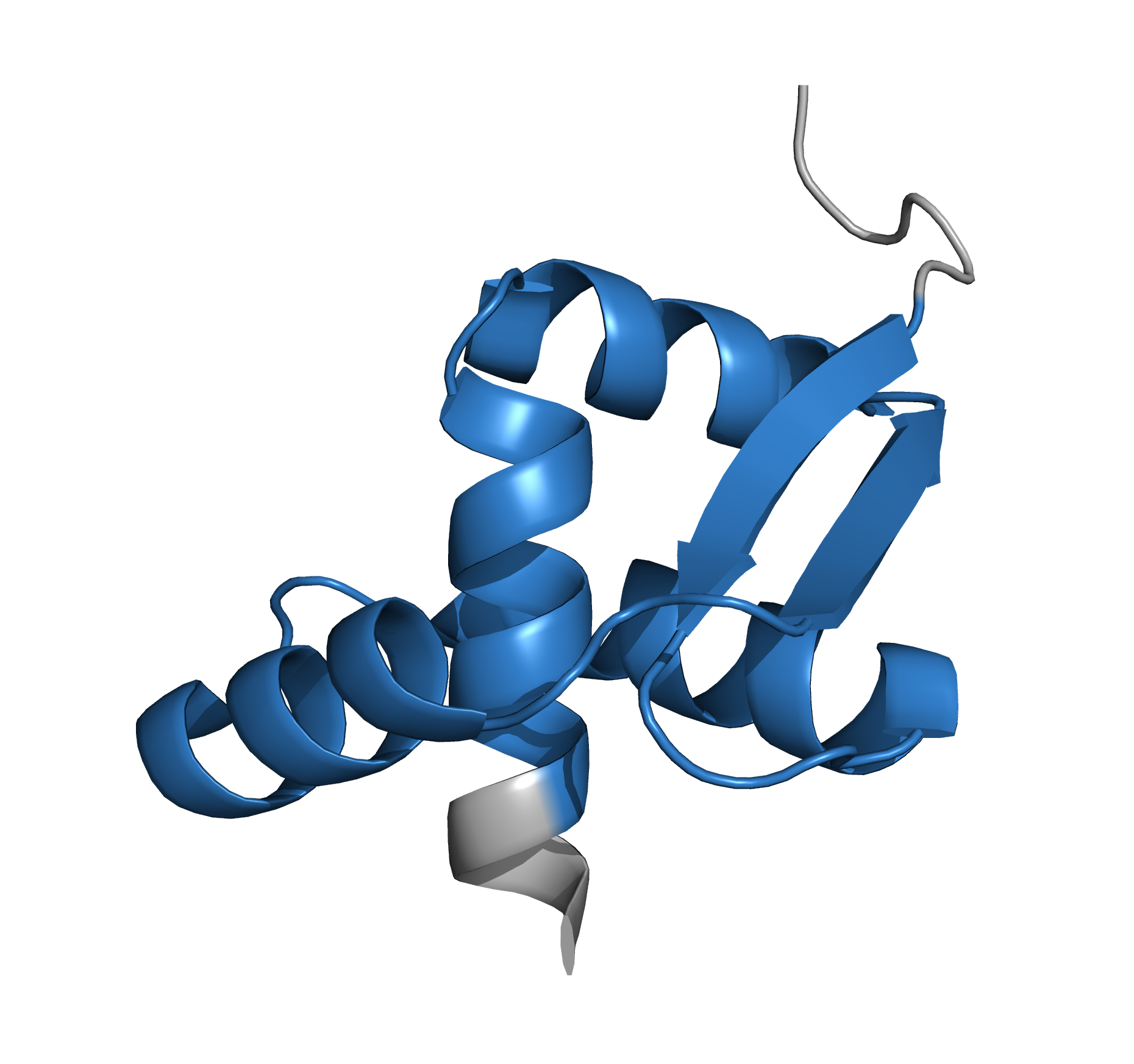}
    \end{tabular}
    
    \caption{3d model and knottines \cite{DabrowskiTumanski2018} of 2EFV. Knot tails are denoted with gray color and not core with blue color.}
    \label{fig:2efv_knot}
\end{figure}

The KnotProt 2.0 package \cite{DabrowskiTumanski2018} detects the knot of type $3_1$ whose core consists of 71 residues, from 6 to 77 amino acids (Figure \ref{fig:2efv_knot}). The knot loop in MJ0366 consists mostly of helices, that probably makes it more tough and thus more stable.  

Although the exact biological function of MJ0366 remains hypothetical, knotted proteins, in general, are known for their mechanical and thermal stability. Some knotted proteins have been implicated in various biological processes, including those related to diseases like Parkinson’s disease.

The study of MJ0366 sheds light on the structural complexity of knotted proteins and their potential stability in extreme conditions.

For calculations model 2EFV from PDB database was used.

\section{Results}
\subsection{Soliton model of MJ0366}

Knotted protein 2EFV consists of 82 amino acids which form 2 $\beta$-strands and 5 $\alpha$-helices. It means that there are 8 loops and the structure must be describe by at least 8 individual solitons. But $\beta$-strands because of their complicated geometry require additional solitons in bending positions. Finally, we construct the native structure of the protein using 12 solitons. It allows us to reach resolution of 0.52 \AA\ RMSD that is less than reported resolution 1.90 \AA\ of experimental data.

\begin{figure}[ht]
    \centering
    \begin{tabular} {c c}
     \includegraphics[width=0.45\linewidth]{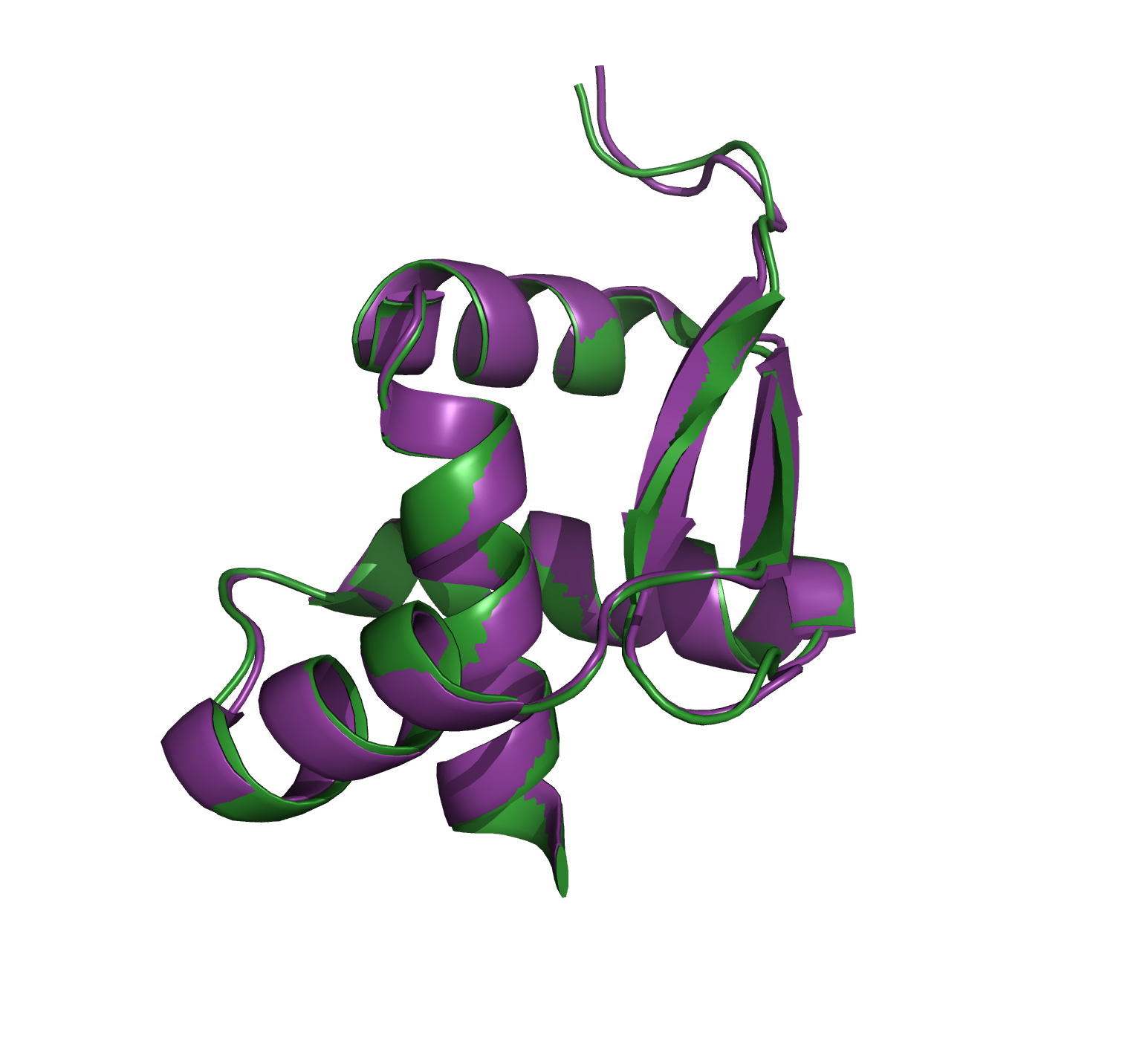}    &  
         \includegraphics[width=0.45\linewidth]{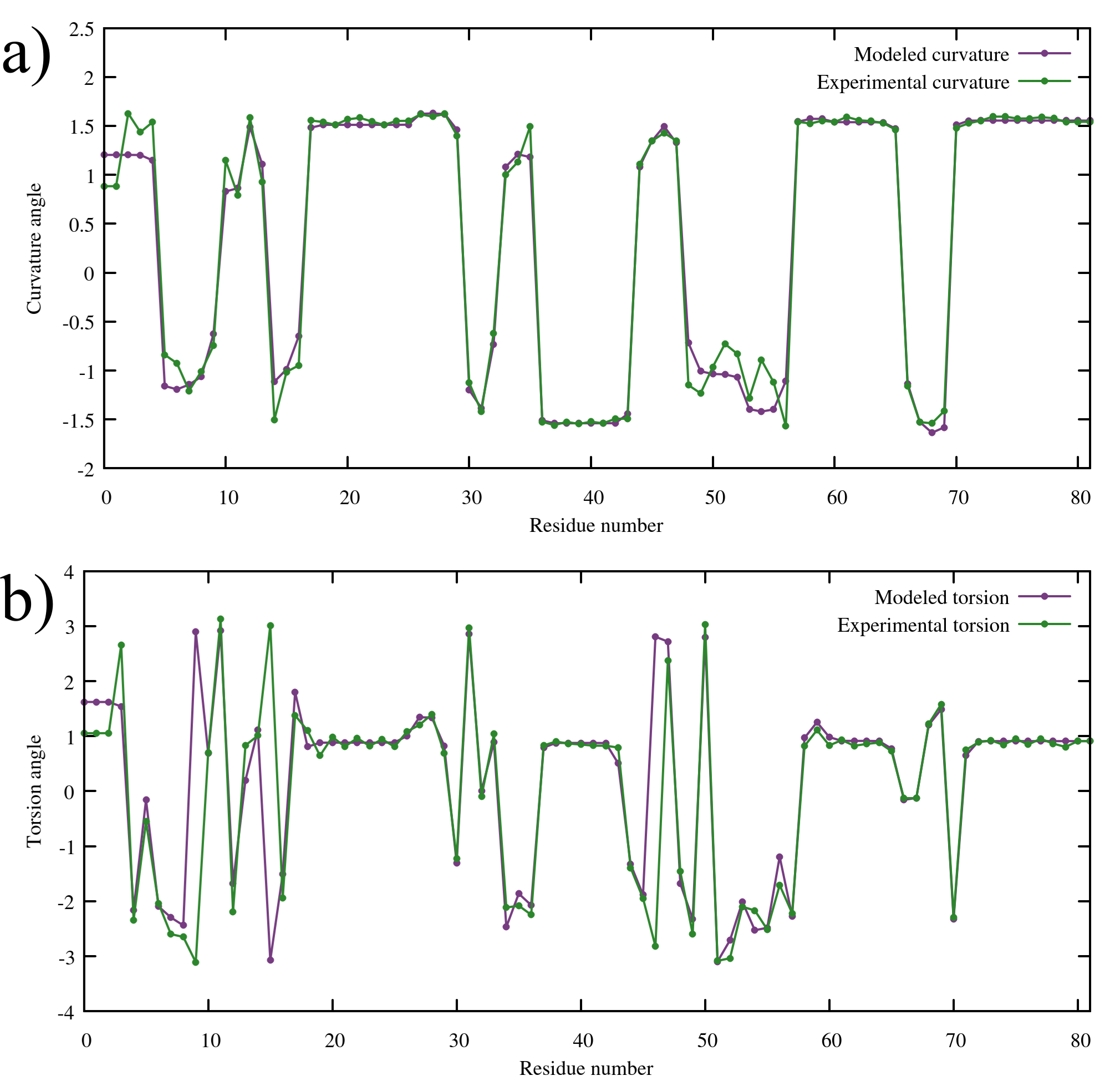}
    \end{tabular}
    
    \caption{Overlaying of experimental model taken from PDB (green) and soliton model (purple). RMSD of them is 0.52 \AA. Right pannel represents comparison of curvature (a) and torsion (b) of soliton model and PDB data.}
    \label{fig:soliton}
\end{figure}

The results of constructing are summarized in Figure \ref{fig:soliton}. In panel a) overlay of 3d models experimental (green) and soliton (purple) structures is represented. Panel b) shows how the angles of curvature and torsion behave in experimental data and in the soliton model.

\subsection{Unfolding-folding simulation results}

We performed simulations of the unfolding-folding process and inspected 3d models of the protein.

\begin{figure}[ht!]
    \centering
    \includegraphics[width=1\linewidth]{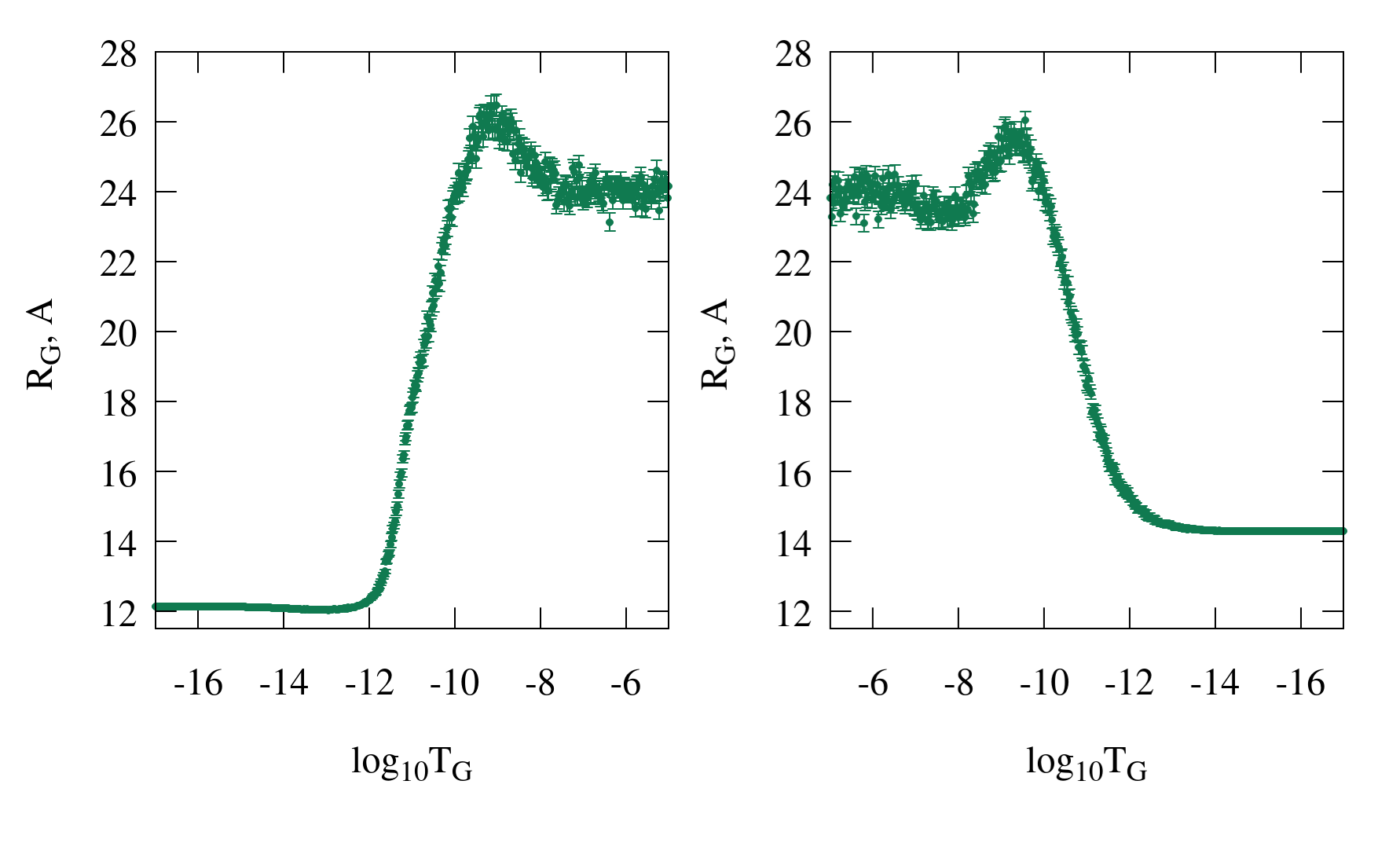}
    \caption{Radius of gyration in heat-cool simulation.}
    \label{fig:R_G}
\end{figure}

We found that 2EFV does not fold back to its native state once knot was untied after slow enough heating to the temperature of self-avoiding random walk (i.e. denaturation). It becomes clear from the graphs of radius of gyration $R_G$ (Figure \ref{fig:R_G}). In the panel c) enlarged area around $T_G=10^{-12}$ preciding to unfolding process is shown. The radius of gyration undergoes a shrinkage that is connected to unknotting of the molecule \cite{Antti2009}. Meanwhile in panel d) which corresponds to the same area, but during the cooling process, shrinkage is totally absent.   

\begin{figure}[ht]
    \centering
    \begin{tabular}{c c}
    \includegraphics[width=0.45\linewidth]{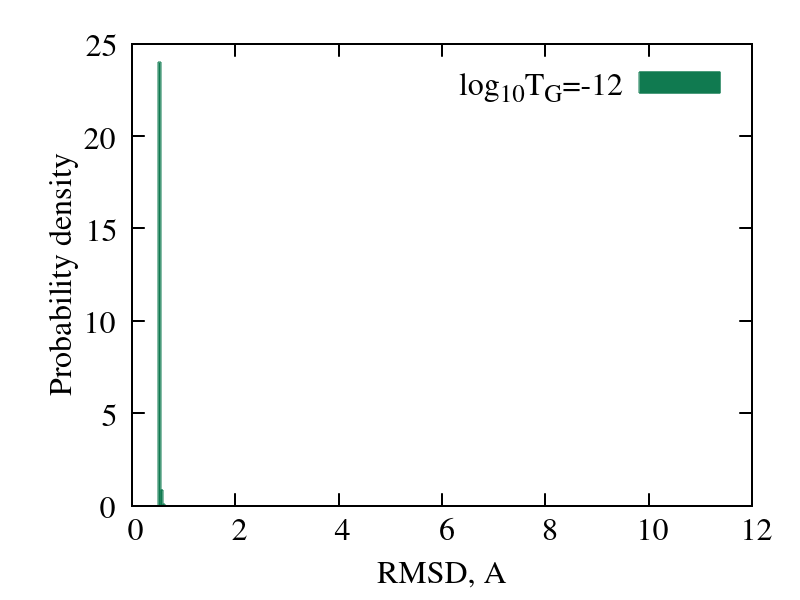}     &  \includegraphics[width=0.45\linewidth]{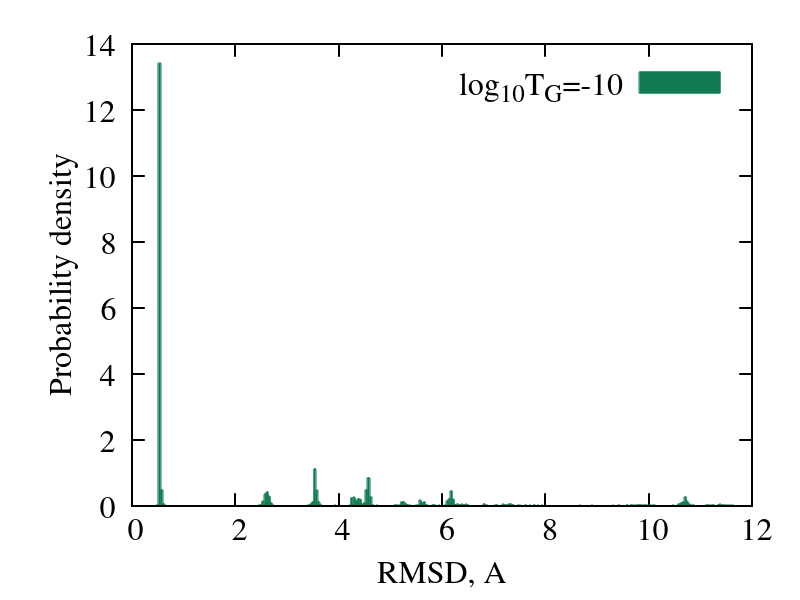} \\
     \includegraphics[width=0.45\linewidth]{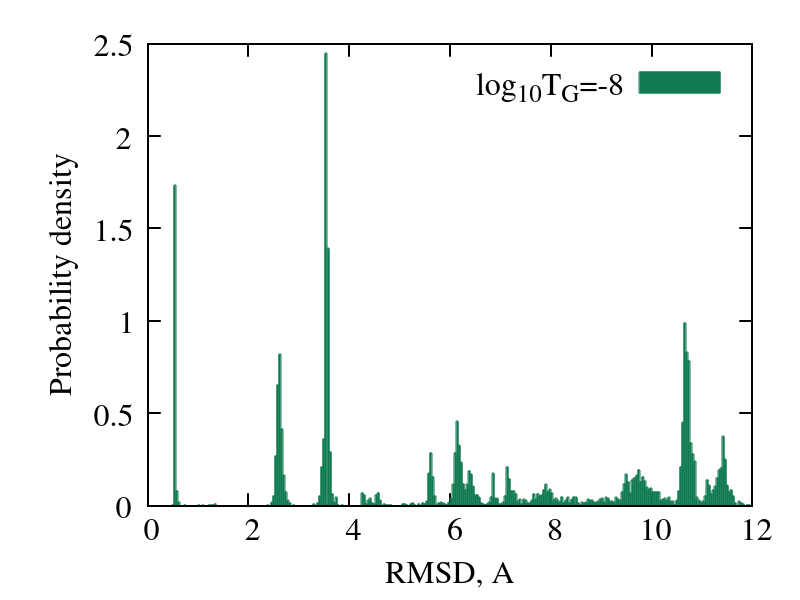}     & \includegraphics[width=0.45\linewidth]{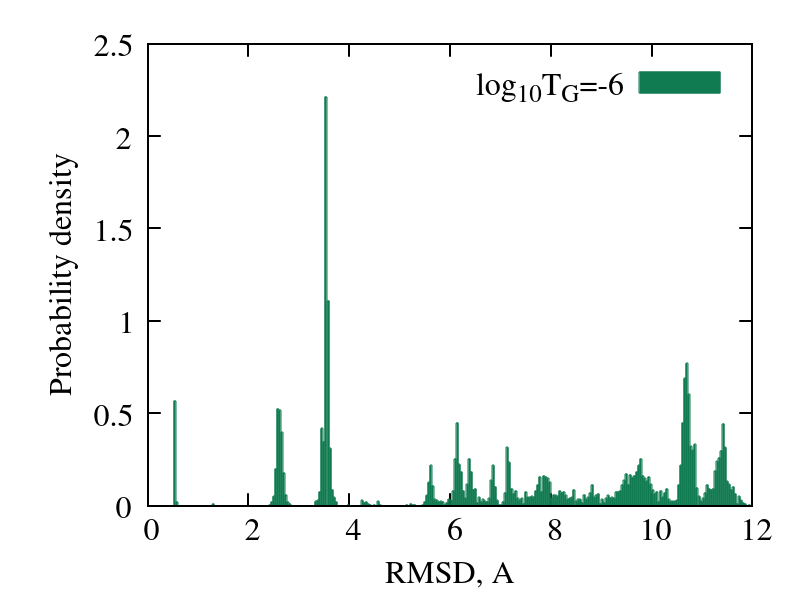} 
    \end{tabular}
    \caption{Probability distribution of RMSD after $N_{steps}=1$ million steps cooling process from certain temperature $T_G$ to temperature of native state $T_{G_0}=10^{-17}$.}
    \label{fig:prob_dist}
\end{figure}

We studied the temperature range in which protein can be folded back to the native state. To do that, we heated the protein to the required temperature ($N_{steps}$), did a short thermalization ($N_{steps}/10$) and then cooled it back to $T_G=10^{-17}$. We present probability distributions of RMSD after heating and cooling to a certain temperature, the results are represented in Figure \ref{fig:prob_dist}.

We checked all obtained configurations by Knot_pull and found that all and only configurations which belong to the first peak must be considered as knotted ones. Other peaks are products of misfolding to local energy minimums. 

\begin{figure}[ht]
    \centering
    \includegraphics[width=1\linewidth]{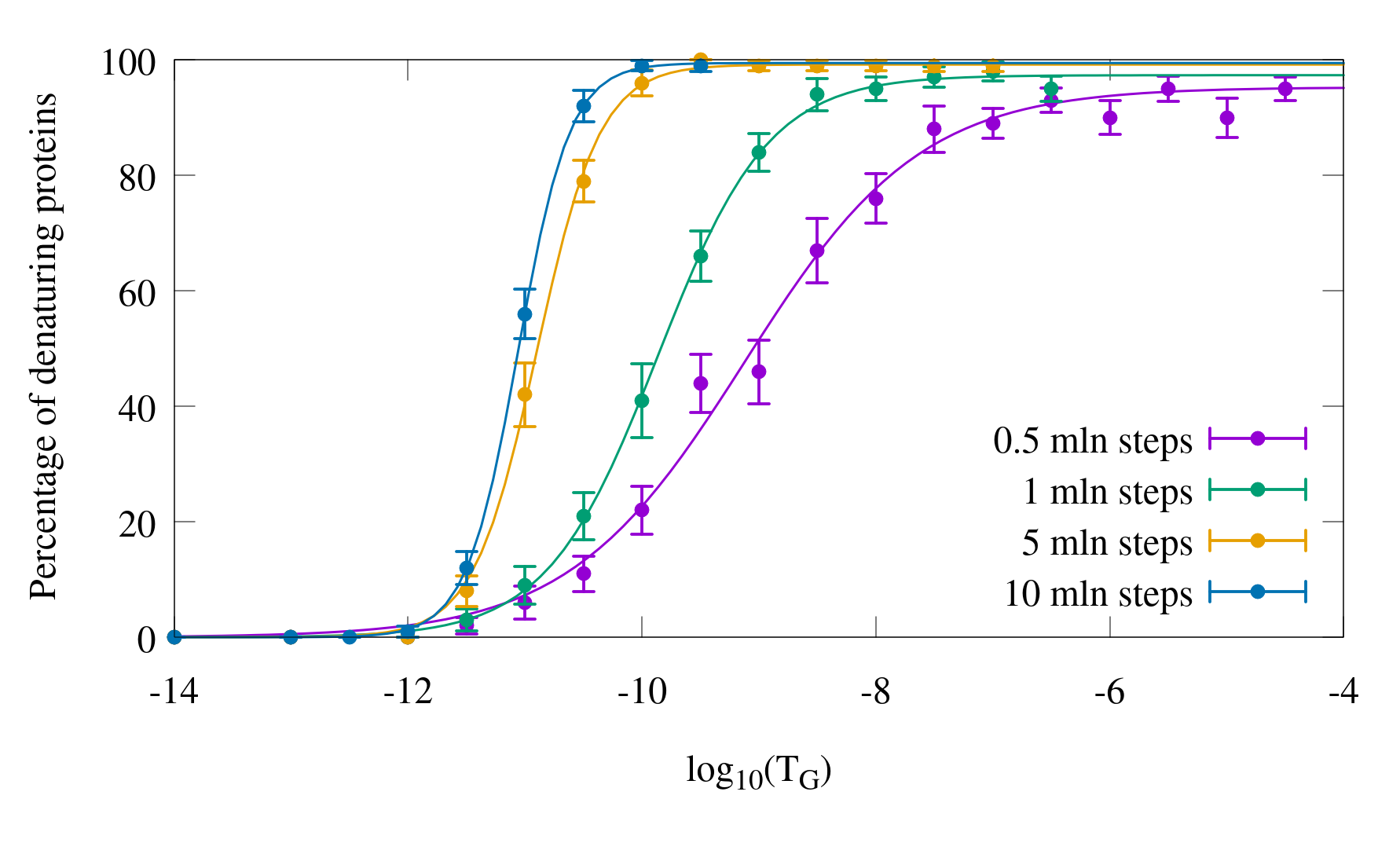}
    \caption{Percent of unknotted configurations}
    \label{fig:unknot}
\end{figure}

The series of simulation was provided. We studied the dependences of the percantage of knotted configurations after heating and cooling from the maximum temperature in the heating-cooling process and the amount of Monte Carlo steps in the process. As follows from trajectories analysis the unknotted configuration can not be tied back, and after the cooling if the configuration is still knotted, it means that it was not untied. So, we can interpret this graph as percent of untied configurations. 

The results obtained were fitted by the sigmoid function 
\beq 
f (T_G) =\frac{a}{(1+e^{-b(\log T_G-\log T_{\mathrm{crit}})})}.
\label{eq:sig}
\eeq

The parameter $T_{\mathrm{crit}}$ in \eqref{eq:sig} can be interpreted as as a critical temperature for particular heating velocity at which protein denatures irreversibly. We plot extracted from fit values of critical temperature and fit it with linear regression. The critical temperature at infinitely slow heating turns out to be 

$$\log_{10}T_{\mathrm{crit}} = -11.18 \pm 0.05. $$

\begin{figure}[ht]
    \centering
    \includegraphics[width=0.8\linewidth]{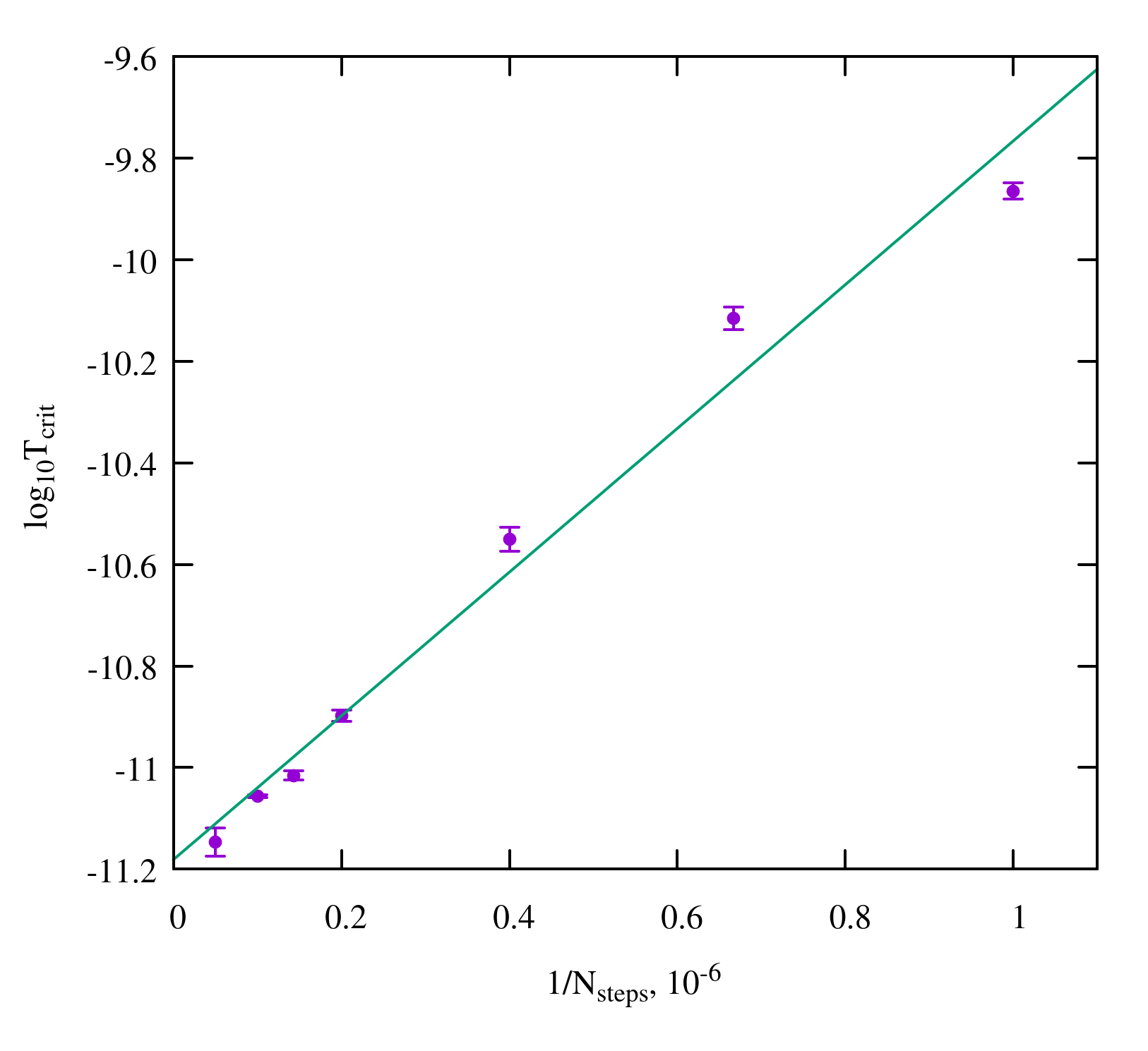}
    \caption{Critical temperature dependence of heating velocity.}
    \label{fig:crit_temp}
\end{figure}


Thus, we see that 2EFV is relatively thermally stable, especially at high heating velocities. We propose that the high thermal stability of 2EFV is connected to the knot in its structure. To prove that we start simulations not from native state but from unknotted deeply cooled configuration. The unknotted configurations we chose from a huge variety of configurations cooled after unknotting. We made several calculations starting from different configurations closest to the native state having RMSD in a range from 3 to 5 \AA\, and radius of gyration from 12.20 to 12.30 \AA.

\begin{figure}[ht]
    \centering
    \begin{tabular}{c c}
    \includegraphics[width=0.45\linewidth]{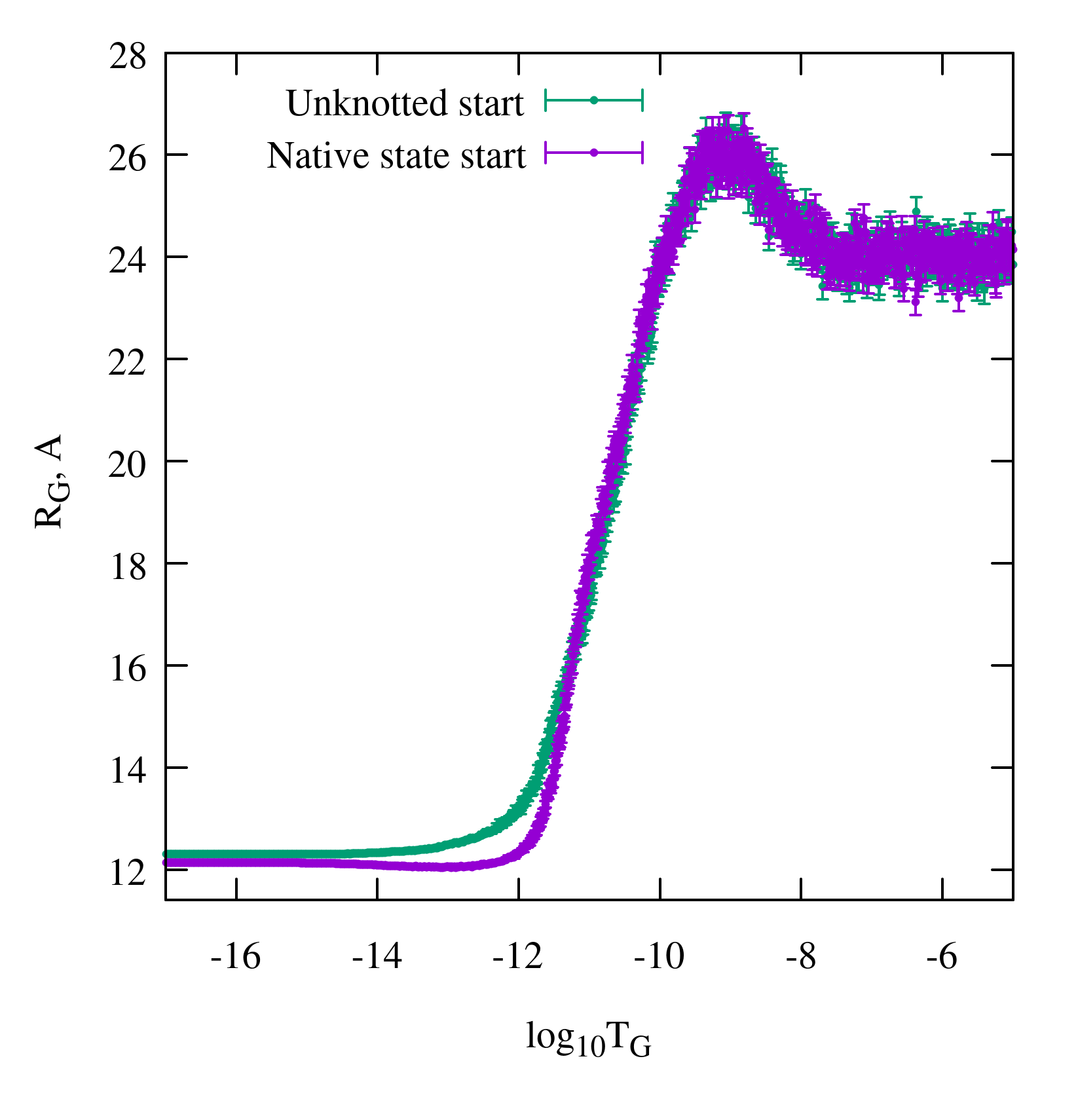}     & 
    \includegraphics[width=0.45\linewidth]{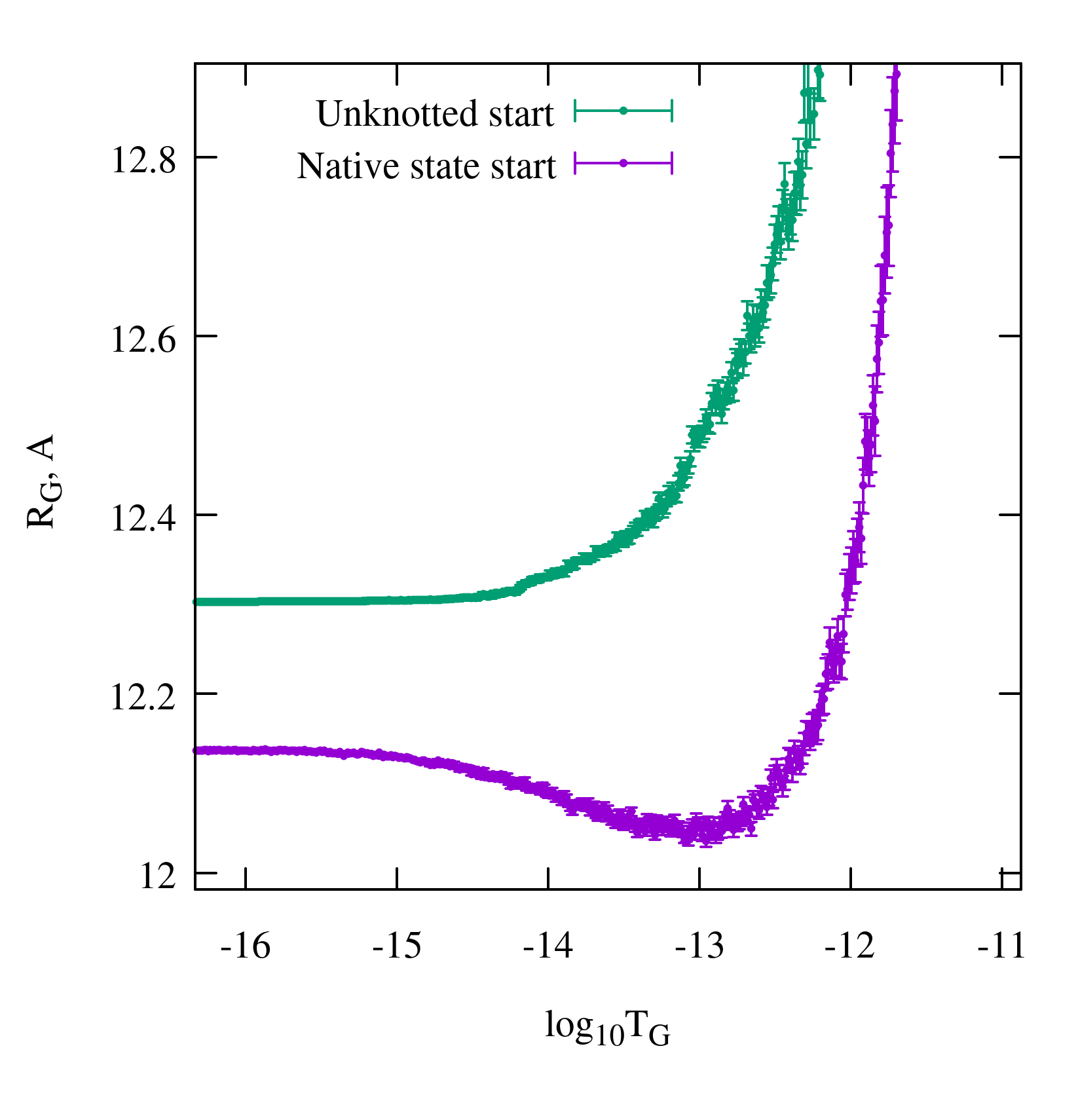}
    \end{tabular}
    
    \caption{Radius of gyration in heat-cool simulation}
\label{fig:knot-unknot}
\end{figure}

It is clear from the graphs that the unknotted configurations starts unfolding earlier for two orders of magnitude of Glauber temperature than the native knotted configuration. Also, it does not shrink, and we can conclude that shrinking before unfolding is connected with the unknotting of the structure. 

\section{Conclusions}
In conclusion, our study demonstrates that topology plays a crucial role in the dynamics of proteins, particularly in their thermal stability and folding behavior. Through the investigation of the protein MJ0366, which contains a trefoil knot in its structure, we have elucidated several key aspects of how this topological feature influences protein behavior.

Our findings indicate that the knot within the MJ0366 protein contributes significantly to its thermal stability. This enhanced stability is intricately linked to the speed of heating, suggesting that the rate of thermal application is a critical factor in maintaining the structural integrity of knotted proteins. During the unfolding process, we observed that the protein undergoes a noticeable shrinking, a phenomenon that correlates with the unknotting of the protein structure. This connection between shrinking and unknotting highlights the mechanical and structural impact of the knot on the protein's response to thermal stress.

Furthermore, our research has shown that once the MJ0366 protein is unknotted, it cannot refold back into its original knotted configuration. This irreversibility underscores the unique and critical role of the knot in maintaining the protein's functional state and stability.

The implications of our study extend beyond the specific case of the MJ0366 protein. Understanding the topology of knots and other forms of self-entanglement in proteins is essential for comprehending their stability, folding mechanisms, and functional properties. As such, future research into the topological characteristics of proteins promises to unveil new insights into protein dynamics, with potential applications in the design of more stable and functional proteins for various biotechnological and medical purposes.

In summary, our work emphasizes the importance of considering topological features in protein studies and opens new avenues for exploring how these features can be harnessed to control and enhance protein stability and functionality.

\section{Acknowlegemnts}

The authors are grateful to Antti Niemi and Alexander Molochkov for useful discussions.
A.B. is supported by the Carl Trygger Foundation Grant No. CTS 18:276. Nordita is supported in part by Nordforsk. AK and AZ are supported by the state assignment of the Ministry of Science and Higher Education of Russia (Project No. FZNS-2024-0002).

\bibliography{2efv_bibl}

\end{document}